\documentclass[10pt,tightenlines,showpacs,aps,prb,twocolumn,amsmath,amssymb,amsfonts,floatfix]{revtex4}
\usepackage{amsxtra}
\usepackage{graphicx}

\def\up{\uparrow}
\def\dwn{\downarrow}

\def\lcc{\langle\!\langle}
\def\rcc{\rangle\!\rangle}
\def\bs{\boldsymbol}
\DeclareMathOperator{\Tr}{Tr}
\DeclareMathOperator{\diag}{diag}

\begin{document}
\title{Full Counting Statistics of Non-Commuting Variables: 
the Case of Spin Counts}
\author{Antonio \surname{Di Lorenzo}\footnote{Current address: 
Department of Physics and Astronomy, Stony Brook
University, SUNY, Stony Brook, NY 11794-3800}}
\author{Gabriele Campagnano}
\author{Yuli V.~Nazarov}
\affiliation{Kavli Institute of Nanoscience Delft, Faculty of Applied Sciences, 
Delft University of Technology, Lorentzweg 1, 2628 CJ Delft The Netherlands} 
\begin{abstract}
We discuss the Full Counting Statistics of non-commuting
variables with the measurement of successive spin counts
in non-collinear directions
taken as an example. We show that owing to an irreducible 
detector back-action,
the FCS in this case may be sensitive to the dynamics of
the detectors, and may differ from the predictions obtained 
with using a naive version of the Projection Postulate.
We present here a general model of detector dynamics and
path-integral approach to the evaluation of FCS.
We concentrate further on a simple ``diffusive" model
of the detector dynamics where the FCS can be evaluated
with transfer-matrix method. The resulting probability 
distribution of spin counts is characterized by anomalously
large higher cumulants and substantially deviates from
Gaussian Statistics.  
\end{abstract}
\maketitle
\section{Introduction}
In the past years, there has been a growing interest 
in noise in mesoscopic systems~\cite{Blanter2000}. 
Normally, noise is an unwanted feature, and, according to classical physics, in principle 
can be made arbitrarily small by lowering the temperature; 
according to quantum physics, however, 
noise is uneliminable due to the intrinsic randomness of elementary processes. Furthermore, 
noise, rather than being a hindrance, contains valuable information 
which adds to the one carried by the mean value of the quantity observed. 
Simple probability distributions, like e.g. the gaussian ones, 
are determined by the mean values and noise. Even though gaussian distributions 
are ubiquitous, there are interesting physical processes which are described 
by non-gaussian distributions. 
Noise alone is not sufficient for the determination of such distributions. 
One needs to know all the momenta, or equivalently their generating function. 
Full Counting Statistics~\cite{Levitov1993,Levitov1996} consists in determining the latter. 

The FCS approach has been receiving increasing attention from the physics community.
Its connection with the formalism of non-equilibrium Green functions~\cite{Keldysh1964} and 
circuit theory~\cite{Nazarov1999} was established~\cite{Belzig2001}.  
It has been used to characterize transport in heterostructures~\cite{Belzig2002}, 
shuttling mechanism~\cite{Pistolesi2004,Romito2004}, charge pumping~\cite{Andreev2001}, 
and multiple Andreev reflections~\cite{Cuevas2003,Johansson2003}. 	 
The technique was extended to charge counts in multiterminal structures~\cite{Nazarov2002}, 
and to spin counts~\cite{DiLorenzo2004}. 
The FCS of a general quantum variable was studied, and the 
necessity of including the dynamics of detectors stressed~\cite{Nazarov2003}. 

There are some open issues in FCS. 
The main one concerns whether it is always possible to find 
 a generating function which allows an interpretation 
in terms of probabilities. 
Indeed, in Ref.~\onlinecite{Belzig2001} 
it was found that such an interpretation is not straightforward. 
This problem was shown to amount to the long standing 
question of the non-positivity 
of the Wigner distribution~\cite{Nazarov2003}. 
The lack of a classical interpretation was 
attributed to the breaking of gauge invariance 
for the charge degrees of freedom, due to the 
presence of superconducting terminals.\cite{Belzig2001} 
It is interesting to consider a more general mechanism 
of gauge invariance breaking 
which involves spin degrees of freedom. 
It may be caused either by the presence of 
ferromagnetic terminals or by subsequent detectors 
measuring different components of 
the spin. We shall consider the latter case.  

Another issue we want to address is the range of applicability of 
the Projection Postulate. Since von Neumann's 
classic work\cite{vonNeumann1932}, 
it is known that Schr\"odinger's evolution cannot account for the fact 
that the result of an individual measurement has a unique value, and 
cannot be described by a superposition. 
It is necessary to supplement Schr\"odinger's evolution with an additional 
evolution (type II in the terminology of von Neumann), projecting 
the state of the observed system into the eigenstate 
of the measured observable corresponding to the actual outcome. 
This can be done at several stages: one could dispense with the description 
of the measurement, and project the wave-function of the system. 
Alternatively, one may continue the chain by describing the interaction 
of system and detector, trace out the system's degrees of freedom, and then 
project the state of the detector. 
This chain can be continued indefinetely, by skipping 
the projection of the detector's state, and considering the coupling 
of the detector with the visible radiation, of the latter with the eye 
of the observer, etc. 
So far, it has been implicitly assumed that the predictions of Quantum 
Mechanics do not depend on the stage at which one chooses to 
stop the chain and project. In this work, we shall demonstrate 
that different statistics are predicted when one projects 
at the level of the system and at the level of the detector. 
The reason for this is that, in the example we shall discuss, 
the quantum dynamics of the detectors cannot be neglected, even 
after accounting for decoherence.

Besides being the simplest illustration of noncommuting variables, 
detection of spin components is a worthy subject in its own right. 
Spintronics, i.e.~the study of how 
producing, detecting, and manipulating spins, 
 is a rapidly growing field~\cite{Awschalom2002}, which has already found 
important technological applications~\cite{Chang2003}. 

In this paper, the subsequent detection of 
non-commuting variables is discussed.  
The Full Counting Statistics approach 
allows to obtain the joint probability distribution 
for the counts. The non-commutativity 
of the observed variables manifests itself in the fact that the 
back-action of detectors, and their quantum dynamics, 
must be taken into account. 
This remains true also when an environment-induced 
dissipative dynamics for detectors is included. 
The reason is that one does not observe one particle 
at a time, but a flux of particles 
traversing the detectors at a rate 
which can be larger than the decoherence rate of the detectors themselves. 
	
The present paper is laid out as follows: 
In section \ref{sec:background} 
the connection of FCS with the density matrix of the detectors is derived, 
and a general theory 
of detection of non-commuting variables is presented; 
a model for the measurement is introduced. 
In section \ref{sec:nodynamics}, 
we discuss the case of ideal quantum detectors, having 
no internal dynamics. We argue that they do not provide a realistic model of 
detectors because of their long memory. 
In section \ref{sec:dynamics}, 
we discuss the internal dynamics of the detectors. 
The fact that detectors are 
``classical'' objects is accounted for by 
introducing a dissipative dynamics due 
to their interaction with an environment. 
Then, since we intend to concentrate on spin counts, 
in section \ref{sec:spindetector}
we present a model for 
a spin detector in solid state, relying on spin-orbit interaction. 
We proceed to section \ref{sec:setup} 
by introducing the particular system that we study, namely 
two normal reservoirs connected by a coherent conductor. 
In section \ref{sec:technique} 
we give details about the derivation of the FCS for this system, relying 
on the full quantum mechanical description of detection process, 
and we present the results. 
In section \ref{sec:PP}, we discuss the 
FCS that would be obtained by a naive application 
of the projection postulate, 
i.e.~by neglecting the quantum dynamics of the detectors. 
In section \ref{sec:comp12}, 
we compare the results of the two approaches for the case 
of one and two detectors in series, and we find that they coincide.  
In section \ref{sec:comp3}, 
we find a discrepancy between the two approaches when 
three detectors in series are considered. 
In particular, we show that both approaches 
predict the same second-order cross correlators, 
and that they differ in the prediction 
of fourth order cumulants $\lcc \sigma_1^2 \sigma_3^2\rcc$. 
Finally, in section \ref{sec:case}, 
the case of three spin detectors, monitoring the $X$, $Y$ and $Z$ 
components of spin current, is presented. 	
The probability distribution for the counts reveals large 
deviations from the Gaussian distribution. 
\section{General considerations about measurements}\label{sec:background}
All the information that we can gain about a system 
is stored in the density matrix of one 
or more detectors (denoted by index $a$) 
which have interacted with the system during a time $\tau$. 
The reduced density matrix is 
\[\Hat{\rho}_{det}(\tau)=
\Tr_{sys}{\left\{ \mathcal{U}_{\tau,0}\Hat{\rho}(0)
\mathcal{U}_{\tau,0}^\dagger\right\}}\;,\]
where $\Tr_{sys}$ stands for the trace over 
the degrees of freedom of the measured system, 
$\Hat{\rho}(0)$ is the initial density matrix of system and detectors, and 
$\mathcal{U}_{\tau,0}$ is the time evolution operator. 
We focus on the representation of $\Hat{\rho}_{det}$ in a basis $|\phi\rangle$, 
$\rho^{\phi,\phi'}_{det}(\tau) \equiv 
\langle\phi|\Hat{\rho}_{det}(\tau)|\phi'\rangle$. 
Here, $|\phi\rangle = \bigotimes_a |\phi_a\rangle$ 
is a vector in the Hilbert space of the detectors. 
Since the time-evolution is linear, a matrix 
$\mathcal{Z}^{\phi,\phi'}_{\mu,\mu'}$ 
exists such that
\begin{align*}
\rho^{\phi,\phi'}_{det}(\tau) 
=&\ \int d\mu d\mu' \mathcal{Z}^{\phi,\phi'}_{\mu,\mu'} 
\rho^{\mu,\mu'}_{det}\!\!(0)\;.
\end{align*} 
Thus, given that one knows the initial density matrix of the detectors, 
$\mathcal{Z}$ contains all the information 
one can extract from the measurement. 
However, part of this information gets lost: we can only know 
the diagonal elements of the density matrix in a particular basis, 
identified by the pointer 
states of the detectors. These states, which will be denoted by
$|N\rangle$, correspond to 
the detectors indicating the values $\{N_a\}$, 
and are individuated by the property that, 
if one prepares the detector in a generic state identified by a density matrix 
$\rho_{det}^{N,N'}$, and then lets the environment act on it, 
the off-diagonal elements of the density matrix in the basis $|N\rangle$ 
will go to zero with an exponential decay. 
We point out that this does not dispense us from invoking a projection at 
some point. The presence of the environment explains how the ensemble 
averaged density matrix reduces to diagonal form in the basis of pointer 
states, but it does not explain how the density matrix of the subensemble 
corresponding to an outcome $N_a$ purifies to the state $|N_a\rangle$. 
This requires invoking the projection postulate for the detector, 
or, equivalently, an evolution dictated by the rules of the 
bayesian approach\cite{Korotkov2002} 
or of the quantum trajectory \cite{Dalibard1992} one.

The quantity accessible to observation is
the probability to find the detectors in states 
$|N_a\rangle$, after a time $\tau$. 
It is given by 
\begin{equation}\label{eq:prob1}
P_\tau(N) = 
\langle N|
\Hat{\rho}_{det}(\tau)
|N\rangle\;.
\end{equation} 
If offdiagonal elements of the detector's density matrix decay instantaneously, 
$P_\tau(N)$ depends only on the probabilities at a time immediately 
preceding $\tau$, $P_{\tau-dt}(N)$, and the process is Markovian.

Now, let us consider the 
operators $\Hat{K}_a$ corresponding to the read-out variables of the 
detectors. Their eigenstates are $|N_a\rangle$, 
where $N_a$ indicates an integer which is proportional to $K_a$. 
The proportionality constant is provided below. 
Let us also introduce the conjugated  
operators $\Hat{V}_a$, 
$\left[\Hat{K}_a,\Hat{V}_b\right]=i\delta_{ab}\hbar$, and their eigenstates 
$|\phi_a\rangle$, 
with $\phi_a$ dimensionless quantitities proportional to $V_a$. 
If we insert to the left and to the right of $\Hat{\rho}_{det}$ 
in the RHS of Eq.~\eqref{eq:prob1}   
the identity (in the detectors' Hilbert space)
in the form 
$\mathcal{I}\propto\int \frac{d\phi}{2\pi} |\phi\rangle\langle \phi|$, 
we obtain 
\begin{align}
\nonumber
P_\tau(N) =&\ \int \frac{d\phi^+}{2\pi}\frac{d\phi^-}{2\pi}\\
&\times \exp{\left[-\frac{i}{\hbar} (\phi^+-\phi^-)\cdot N\right]} 
{\rho}^{\phi^-,\phi^+}_{det}(\tau)\;.
\end{align}
We used the shorthand $\phi\cdot N \equiv \sum_a \phi_a N_a$. 
We change variables according to $\phi^\pm = (\Phi \pm \phi)/2$. 
Here, $\Phi$ and $\phi$ 
are the classical and quantum part of the field, respectively. 
This terminology reflects the fact that fluctuations of $\Phi$ 
are set by the temperature, while fluctuations of $\phi$ depend on 
$\hbar$, as we shall prove in section \ref{sec:dynamics}.   
The time evolution depends on the Hamiltonians of the system and the detectors, 
and on their interaction. 
We focus on the detection of internal degrees of freedom 
of a system whose center of mass coordinate 
$\mathbf{x}$ is not affected by the presence of the detectors. 
We consider several detectors in series along the path $\mathbf{x}(t)$. 
We take the interaction to be of the form $H_{int} = \sum_a H_{int}^a$ with 
\begin{equation}\label{eq:int}
H_{int}^a=-\alpha_a(\mathbf{x})\lambda_a \Hat{V}_a \Hat{J}_a\;,
\end{equation}
where $\mathbf{x}$ is the coordinate of the wave-packet, 
$\lambda_a$ coupling constants depending 
on the actual detection setup, 
$\Hat{J}_a$ is an operator 
on the Hilbert space of the system's degrees of freedom, and 
$\alpha_a(\mathbf{x})$ is a function which  
is unity inside the sensible area of the $a$-th detector and zero outside.  
For a one-dimensional motion, e.g., we would have 
$\alpha(x)=\theta(x-X^{(in)})\theta(X^{(fin)}-x)$, with 
$X^{(in)}$ and $X^{(fin)}$ are the coordinates 
delimiting the sensible area of the detector, 
$\theta(x)=0$ if $x< 0$, $\theta(x)=1$ if $x\ge 0$. 
$\Hat{J}_a$ is the current associated with the measured quantity, such that 
the output of the detector does not depend 
on the time each particle takes to cross 
its sensible area. 
Indeed, the equation of motion for the ``measuring'' operator is 
\[\frac{d\Hat{K}_a(t)}{dt} = \alpha_a(\mathbf{x}(t))\lambda\Hat{J}_a(t)\;.\]
In the equation above, we have assumed that 
the operator $\Hat{K}_a$ commutes with 
the unperturbed Hamiltonian of the detector. 
In general, however, $<\Hat{K}_a>$ will fluctuate in time 
due to background noise. Such fluctuations put a lower 
limit to the resolution of the detector.  
For a reliable detection, the resolution must be smaller than 
the minimal variation $K_{Qa}$ one intends to measure. 

Let us introduce proper units. 
We consider the case where the measured 
quantities have discrete values proportional to a quantum $E_{Qa}$. 
For instance, for charge $E_Q=e$, the elementary charge, and for spin 
$E_Q=\hbar/2$. Every time an elementary unit passes the detector, 
the readout of the latter will change by $K_{Qa}= \lambda_a E_{Qa}$. 
Thus, we introduce the number and phase operators 
$N_a = K_a/K_{Qa}$, $\phi_a = V/V_{Qa}$, 
with $V_{Qa}=\hbar/K_{Qa}$. 
We further assume that 
\emph{i}) the detectors are initially prepared in a state with zero counts 
$\Hat{\rho}_{det}(0)=|N\!\!=\!\!0\rangle\langle N\!\!=\!\!0\,|$, 
and \emph{ii}) the spread of the system wave-packet is much smaller than  
the distance between two subsequent detectors, 
$\Delta x \ll X_{a+1}-X_a$. 
The first assumption implies that
\[\rho^{\phi,\phi'}_{det}(\tau) = 
\mathcal{Z}(\phi,\phi')\equiv
\int \frac{d\mu}{2\pi} \frac{d\mu'}{2\pi} 
\mathcal{Z}_{\mu,\mu'}^{\phi,\phi'}\;,
\]
or, explicitly, 
\begin{align}
\nonumber
\mathcal{Z}(\phi^+,\phi^-) =&\ 
\int \frac{d\mu^+}{2\pi}\frac{d\mu^-}{2\pi} 
\int\limits_{\mu^+}^{\phi^+}\!\mathcal{D}\phi^+(t) 
\!\int\limits_{\mu^-}^{\phi^-} 
\!\mathcal{D}\phi^-(t) \\
&\label{eq:FCS_definition}
\exp{\left(S_{det}[\phi^+]-S_{det}[\phi^-]+
\mathcal{F}_{sys}[\phi^+,\phi^-]\right)},
\end{align}
where 
the limits of the path-integrals fix the values of the fields at 
$t=0$ and $t=\tau$, and 
we introduced the influence functional of the system 
on the detectors\ \cite{Feynman1963a}
\begin{align}\nonumber
&\exp{
\left(\mathcal{F}_{sys}[\phi^+,\phi^-]\right)}:=
\Tr_{sys} \\
&{\left\{
\exp{\left(S_{int}[\phi^+,\Hat{J}]\right)}
\Hat{\rho}_{sys}(0)
\exp{\left(-S_{int}[\phi^-,\Hat{J}]\right)} 
\right\}},
\end{align}
where $S_{int}$ is the action corresponding to the interaction 
$H_{int}$ given in Eq.\ \eqref{eq:int}.  
We shall call $\mathcal{Z}$ the quantum generating function. 
In principle it depends on twice as many parameters 
than the classical generating function does. 
In the rest of the paper we shall use the cumulant generating function (CGF), 
$\mathcal{F}\equiv \log{\mathcal{Z}}$ .  
The advantage of working with the CGF is that it often has 
a clearer interpretation than $P_\tau$, since independent processes 
contribute factors to $P_\tau$ and simply additive terms to the CGF. Hence, 
if subsequent events are independent, the CGF is proportional 
to the observation time $\tau$. 
Thus, time averaged cumulants, which for long $\tau$ correspond 
to zero-frequency noise and higher order correlators for currents, 
have a finite value. 
\section{Detectors with no dynamics}\label{sec:nodynamics}
We analyze the situation where the dynamics of the detectors is neglected. 
This means that 
\[
\exp{S_{det}[\phi(t)]}=\prod_t \delta(\phi(t)-\phi),\]
i.e.\ the counting fields are constant. 

We consider first the case of one detector. 
Then 
$\mathcal{Z}^{\phi^+,\phi^-}_{\mu^+,\mu^-} = \delta_{\phi^+,\mu^+} 
\delta_{\phi^-,\mu^-} \mathcal{Z}(\phi^+,\phi^-)$, and 
\begin{align}
\mathcal{Z}^{\phi^+,\phi^-} =&
\Tr_{sys}{\left\{ 
\mathcal{U}^{\phi^+} 
\Hat{\rho}_{sys}(0){\mathcal{U}^{\phi^-}}^\dagger\right\}}
,
\end{align}
where
\begin{align*}
\mathcal{U}^{\phi} =&\  \mathcal{T}
\exp{\left[- i \phi\int dt  \Hat{J}(t)/E_Q\right]} 
\end{align*}
($\mathcal{T}$ being the time-ordering operator) 
is an operator in the system's Hilbert space. 
By exploiting 
the cyclic property of the trace, we have that, 
if $\Hat{J}$ is a conserved operator or, 
more generally, $[\Hat{J}(t),\Hat{J}(t')]=0$, 
then $\mathcal{Z}(\phi^+,\phi^-)$
\emph{depends only on $\phi=\phi^+-\phi^-$}. 
It has been shown that in this case 
$\mathcal{Z}(\phi^-,\phi^+)$ gives directly 
the generating function~\cite{Nazarov2003}. 

Next, we consider the case of two detectors. 
The kernel $\mathcal{Z}$ is now 
\begin{align}
\label{eq:gftwodet}
\mathcal{Z}(\phi^+,\phi^-) =&
\Tr_{sys}{\left\{ 
\mathcal{U}^{\phi_2^+} \mathcal{U}^{\phi_1^+}
\Hat{\rho}_{sys}(0)
{\mathcal{U}^{\phi_1^-}}^\dagger{\mathcal{U}^{\phi_2^-}}^\dagger\right\}}
\;.
\end{align}
Here we exploited assumption \emph{ii}), 
and defined 
\begin{align*}
\mathcal{U}^{\phi_a} =&\  
\mathcal{T}
\exp{\left[- i \phi_a\int dt  \Hat{J}_a(t)/E_Q\right]}\;.
\end{align*}
Once again we exploit the cyclic property of the trace 
and see that the expression 
does not depend on the combination 
${\Phi}_2\equiv \phi_2^++\phi_2^-$. 
From Eq.~\eqref{eq:gftwodet} we see that in general, for two detectors, 
$\mathcal{Z}$ does depend on $\Phi_1$, even when $\Hat{J}_a$ are conserved. 
However, when the system is initially in the unpolarized state 
$\Hat{\rho}_{sys}\propto \mathcal{I}_{sys}$, 
the dependence on ${\Phi_1}$ disappears as well. 
Another case in which this happens is when 
the detectors monitor two commuting degrees 
of freedom which are conserved. For instance, 
if the current $\Hat{J}$ is not conserved, 
in general $\left[\Hat{J}(t),\Hat{J}(t')\right]\neq 0$. 
Thus, even if one repeats the same measurement, 
one would obtain different results. 
If however the current is conserved $\Hat{J}(t)=\Hat{J}$, 
and both detectors 
measure $\Hat{J}$, the kernel depends only on the combination $\phi_1+\phi_2$, 
which means that the two measurements will give the same result.  

In general, when there are three detectors, 
labelled 1, 2 and 3 according to their order, 
measuring non-commuting quantities, even if the 
system is initially unpolarized, 
the integrand will depend on the classical variable of the middle 
detector, $\Phi_2$. 
When such a dependence appears in the expression 
for the generating function, 
it is a signal that the internal dynamics 
of the detector must be taken into account. 
Indeed, when $\mathcal{Z}$ does not depend on $\Phi$, 
the density matrix is diagonal 
in the basis $|N\rangle$. 
When $\mathcal{Z}$ does depend on $\Phi$, 
$\rho_{det}$ develops off-diagonal components. 
We consider as an example the case where the detectors' density matrix is 
prepared in a diagonal state at $t=0$, and two particles are sent 
to the detectors one at time $t_1>0$ and the other at time $t_2>t_1$, 
in such a way that their wave-packets do not overlap. 
Then, after the first particle has crossed the detectors, 
the density matrix of the detectors 
$\rho_{det}^{N,N'}$ has offdiagonal elements, 
which depend on the original diagonal elements (probabilities). 
Since one observes only the probabilities, 
this can not be ascertained directly. 
However, when the second particle crosses the detectors, the new probabilities 
will be a combination of the former diagonal and off-diagonal elements. 
In order to know $\rho_{det}^{N,N}(t_2)$, knowledge of 
$\rho_{det}^{N,N}(t_1)$ is not sufficient.  
Thus, the process is non-Markovian. 
In principle, even after the detector 
has been measuring for a long time a large number of particles, 
the off-diagonal 
elements created after the passage of the first particle 
 will still influence its dynamics. 
This is not realistic, since, 
because of the coupling of the detectors to the environment, 
the offdiagonal elements will go to zero within a typical time $\tau_c$. 
In order to account for this, one should 
consider the dynamics of the detectors, 
which we shall do in the next section. 

\section{Detectors with internal dynamics}\label{sec:dynamics}
We model the decoherence 
of the detectors by introducing a dissipative 
dynamics for the detectors' degrees of freedom, i.e. 
we couple the detectors to an environment, whose degrees of 
freedom are traced out. 
We model the environment as a system of independent harmonic oscillator 
in thermal equilibrium, having the action 
\[S_{env} = -\frac{i}{\hbar}\int dt \sum_j 
\frac{1}{2} m_j\left[ \dot{x}_j^2 - \omega_j^2 x_j^2\right]
,\]
and coupling to the detectors through the position operator 
\begin{equation}\label{eq:det-env}
S_{det-env}=\frac{i}{\hbar} \int dt \sum_{ja} c_{ja} x_j V_{Qa} \phi_a,
\end{equation}
with $c_{ja}$ coupling constant between the $j$-th oscillator 
and the $a$-th detector. 
Then the generating function becomes
\begin{widetext}
\begin{align}\nonumber
&\mathcal{Z}(\phi^+,\phi^-) =
\int \frac{d\mu^+}{2\pi} \frac{d\mu^-}{2\pi} \int dx_j dx^+_j dx^-_j
\int\limits_{\mu^+}^{\phi^+}\!\mathcal{D}\phi^+(t) 
\!\int\limits_{\mu^-}^{\phi^-} 
\!\mathcal{D}\phi^-(t)
\int\limits_{x_j^+}^{x_j}\!\mathcal{D}x^+_j(t)
\int\limits_{x_j^-}^{x_j}
\!\mathcal{D}x^-_j(t)\rho_{env}(x^+,x^-)\\
&\exp\biggl\{
S_{det}[\phi^+]-S_{det}[\phi^-]+F_{sys}[\phi^+,\phi^-]+
S_{env}[x^+_j]-S_{env}[x^-_j] 
+ S_{det-env}[x^+_j,\phi^+]-S_{det-env}[x^-_j,\phi^-]
\biggr\}
\;,
\end{align}
\end{widetext}
In the expression above, we isolate the part 
\begin{align}
\nonumber
&\exp{\mathcal{F}_{env}} = \int dx_j dx^+_j dx^-_j
\int\limits_{x_j^+}^{x_j}\!\mathcal{D}x^+_j(t)
\int\limits_{x_j^-}^{x_j}
\!\mathcal{D}x^-_j(t)\\
&\nonumber
\exp\biggl(S_{env}[x^+_j]-S_{env}[x^-_j] +S_{det-env}[x^+_j,\phi^+]\\
&\phantom{S_{det-env}}-S_{det-env}[x^-_j,\phi^-]\biggr)
\rho_{env}(x^+,x^-)
\end{align}
which gives the influence functional of the environment on the detectors. 
We notice from Eq.\ \eqref{eq:det-env} that, 
since the functions $\phi^\pm_a(t)$ are fixed by the external path-integrals, 
they act as 
an external source $I^\pm_j(t)=\sum_a c_{ja} \phi^\pm_a(t)$ on the 
$j$-th harmonic oscillator. 
It is then possible to perform the independent gaussian path-integrals 
over $x_j$, resulting in\cite{Kleinert2004}
\begin{align}
\nonumber
\mathcal{F}_{env} =&\ -\frac{i}{\hbar}
\sum_a V_{Qa}^2\int_{0}^{\tau} dt \int_{0}^{t} dt'
\left(\phi_a^+(t)-\phi_a^-(t)\right)
 \\
&\times\left[\alpha_a(t-t')\phi_a^+(t')
-\alpha_a^*(t-t') \phi_a^-(t')\right]\;,
\end{align}
where the influence of the environment is contained in the 
complex functions $\alpha_a(t)$, whose Fourier transforms are 
\begin{align}
\alpha_a(\omega)=&\ \frac{1}{2}\left(\coth{\frac{\hbar\beta\omega}{2}}+1\right) 
\sigma_a(\omega)\;,\end{align}
where the inverse temperature $\beta=1/k_B T$ comes from 
having assumed the bath in thermal 
equilibrium ($\Hat{\rho}_{env} = \exp{(-\beta \Hat{H}_{env})}$), 
and $\sigma_a$ are the spectral densities
\begin{align}
\sigma_a(\omega)=&\ \pi \sum_j \frac{c_{ja}^2}{m_j\omega_j} 
\left[\delta(\omega-\omega_j)-\delta(\omega+\omega_j)\right]\;,
\end{align}

At low frequencies, 
we can approximate the odd-functions $\sigma_a$ by
$\sigma_a(\omega)\simeq \gamma_a \omega$ (Ohmic approximation), with 
$\gamma_a$ friction constant, as will be clear later. 
We introduce new variables $\phi=\phi^+-\phi^-$, $\Phi=\phi^++\phi^-$.
Thus we get 
\begin{align}
\nonumber
\mathcal{F}_{env} =&\ \sum_a \gamma_a V_{Qa}^2\biggl\{ 
\frac{1}{2\hbar} \int \frac{d\omega}{2\pi} 
\omega \Phi_a(\omega)\phi_a(-\omega)
\\
&-\frac{1}{\beta\hbar^2} 
\int \frac{d\omega}{2\pi}  
\frac{\beta\hbar\omega}{2}\coth{\frac{\beta\hbar\omega}{2}}
\left|\phi_a(\omega)\right|^2
\end{align} 
We take the action of free detectors to be that of harmonic oscillators, i.e.~ 
\begin{equation}
\mathcal{S}_{det}[\phi]=\sum_a 
\frac{-i m_a V_{Qa}^2}{2\hbar}\int \frac{d\omega}{2\pi} 
 (\omega^2-\Omega_a^2) \left|\phi_a(\omega)\right|^2\;,
\end{equation}
where $m_a$ is the ``mass" of the detector (i.e. it is the inertial term 
corresponding to the kinetic energy $m_a V_{Qa}^2\dot{\phi}_a^2/2$). 
Then the generating function reads 
\begin{align}
\nonumber
&\mathcal{Z}(\phi,\Phi) = \int \frac{d\mu}{2\pi} \frac{dM}{2\pi} 
\int\limits_{\mu}^{\phi}\!\mathcal{D}\phi(t) 
\!\int\limits_{M}^{\Phi} 
\!\mathcal{D}\Phi(t)\\
&\nonumber
\exp\biggl\{\sum_a\biggl[
\frac{-i m_a V_{Qa}^2}{2\hbar} \int \frac{d\omega}{2\pi}
\Phi_a(\omega) g_a^{-1}(\omega) \phi_a(-\omega)
\\
&
-\frac{\gamma_a}{\beta\hbar^2}\int \frac{d\omega}{2\pi}  
f(\omega)\left|\phi_a(\omega)\right|^2\biggr]
+\mathcal{F}_{sys}[\phi,\Phi]\biggr\}
\;,
\end{align}
where we introduced 
the response function,
\begin{equation} \label{eq:response}
g_a^{-1}(\omega)=\omega^2-\Omega_a^2
+i\frac{\gamma_a}{m_a} \omega\;,
\end{equation}
from which one can see that $\gamma_a$ are proportional to the friction 
constant, 
and the fluctuation term 
\begin{equation}\label{eq:fluctuation}
f(\omega)=\frac{\beta\hbar\omega}{2}\coth{\frac{\beta\hbar\omega}{2}}
.\end{equation} 
The part of the action containing the fluctuation term in $\phi(\omega)$ 
is, at low frequencies, proportional to temperature $T$ and to $1/\hbar^2$.
 The factor $1/\hbar^2$ strongly suppresses large fluctuations in $\phi$. 
Thus, the influence functional due to the measured system 
$\mathcal{F}_{sys}[\Phi,\phi] = \int dt L_{inf}(\Phi(t),\phi(t))$ 
can be approximated by 
$\int dt L_{inf}(\Phi(t),\phi):= \mathcal{F}_{\phi}[\Phi]$. 
Integration over $\phi_\omega$ gives finally
\begin{align}
\mathcal{Z}(\phi,\Phi) 
=& 
\int dM 
\int\limits_{M}^{\Phi}
\!\mathcal{D}\Phi(t)\  
e^{\mathcal{F}_{\phi}[\Phi]+S_{eff}[\Phi]}\;,
\end{align}
with the effective action 
\[S_{eff}[\Phi]= -\frac{1}{2}
\sum_a \frac{(\beta m_a V^2_{Qa})^2}{\gamma_a} \int d\omega  
\frac{|g^{-1}_a(\omega)|^2}{f(\omega)} \left|\Phi_{a}(\omega)\right|^2 \;.
\]
We notice that at high temperatures $f(\omega)\simeq 1$, and thus 
 $\hbar$ disappears in the 
effective action for $\Phi$. 
For this reason the latter is termed the ``classical'' part of the field. 

In the limit of small mass $m_a\to 0$, 
$m_a\Omega_a^2 V_{Qa}^2\to E_a$, where $E_a$ has a finite value and 
is a typical energy scale of detector $a$,
the effective action simplifies to 
\[S_{eff}[\Phi]= -\frac{1}{2}\sum_a\int dt \left[
\tau_{ac} \left(\dot{\Phi}_{a}(t)\right)^2 
+ \frac{1}{\tau_{ac} \Delta \Phi_a^2}\Phi_a(t)^2\right] 
\;,
\]
with $\tau_{ac} = {\beta\gamma_a V_{Qa}^2}/{2}$ 
the ``coherence time'' of the detector, 
and $\Delta \Phi_a = 2/\beta E_a$ the spread of $\Phi_a$. 
\section{Spin detector}\label{sec:spindetector}
\begin{figure}[t!]
\includegraphics[width=0.4\textwidth]{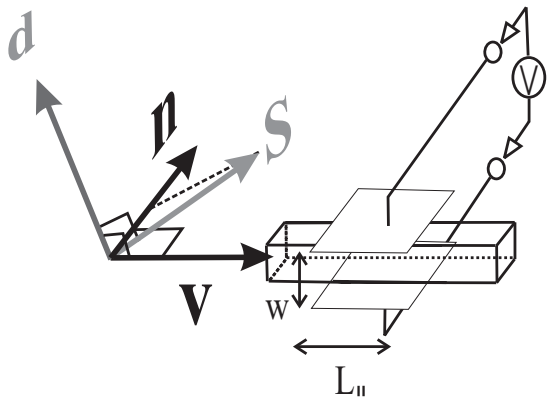}
\caption{\label{fig:detector}
The proposed spin current detector. An electron with velocity $\mathbf{v}$ 
and spin $\mathbf{S}$ induces a voltage drop in a capacitor. The electric 
field $\mathbf{E}$ inside the capacitor produces an Aharonov-Casher phase shift 
on the electrons.}
\end{figure}

We discuss a model for spin detection. 
The setup corresponds to the one proposed and used 
in~\cite{Cimmino1989} to detect Aharonov-Casher effect~\cite{Aharonov1984} 
for neutrons. 
This setup exploits the fact that 
a moving magnetic dipole generates an electric one~\cite{Costa1967,Fisher1971}. 
To measure this, one encloses  
the two-dimensional current lead between the plates of a 
capacitor as shown in Fig.~\ref{fig:detector}. 
While in Ref.~\onlinecite{Cimmino1989} the neutrons 
passed a fixed electric field, which gave 
a constant Aharonov-Casher phase, 
in a spin detector the initial voltage applied to the 
plates is zero, and the passing of a particle with spin 1/2 
will cause the charge 
in the capacitor to 
show pulses towards positive or negative values depending on the result of the 
measurement. The associated phase $K_t = \int_0^t dt Q(t)$ will thus 
increase or decrease stepwise in the 
ideal situation where spins are transmitted 
separately in vacuum through the detector. 

Each spin moving with
velocity $\mathbf{v}$ produces an electric field.
For electrons in vacuum, 
the interaction term between spin and detector 
is given by the spin-orbit coupling 
\[H_{int} = - \frac{1}{2}\bs{E} \cdot (\frac{\bs{v}}{c^2}\times \bs{\mu})\;,\]
with $c$ the speed of light, and the factor $1/2$ 
accounts for the Thomas precession.   
The magnetic moment $\bs{\mu}$ is proportional to the spin  
$\bs{\mu}= (g_S |e|/2m_e) \bs{S}$, 
with $m_e$ mass of the electron, $e=-|e|$ its charge, and $g_S$ its 
spin gyromagnetic factor. 
Thus, we rewrite the interaction as 
\[H_{int} = - (g_S |e|/4m_e c^2) \bs{E} \cdot (\bs{v}\times \bs{S})\;.\]
The spin-orbit coupling induces a current in the $RC$ circuit. 
The integrated charge traversing the circuit is the detector read-out. 
The read-out signal is 
proportional to spin current
in the lead $\mathbf J$, 
$Q=\lambda \mathbf{n}\cdot\mathbf{J}$, 
$\mathbf{n}$ being the unit
vector perpendicular to the direction of the current flow
and parallel to the plates of the capacitor, 
$\lambda$ being a proportionality coefficient. The concrete expression 
for the latter,
$\lambda=g_S |e| L_\shortparallel/4m_e c^2 w$,
depends on the geometrical dimensions of the detector
the length of its plates in the direction 
of the current $L_\shortparallel$, and the distance between the plates $w$. 
The variable canonically conjugated to the read-out is the
voltage $V$ across the capacitor, and the expression for the
interaction in terms of $V$
contains the same proportionality coefficient $\lambda$, 
$H_{int}=-\lambda V \mathbf{n}\cdot\mathbf{J}$. 
Our choice of the detection setup is motivated by the fact
that this detector does not influence electron transfers through
the contact and only gives 
the minimal feedback compatible with the uncertainty principle:
the electrons passing the capacitor in the direction of current
acquire an Aharonov-Casher phase shift, 
which consists in a precession of the spin around 
the detecton axis $\mathbf{n}$. This
depends on spin and is given by 
$\Phi_{AC}= \lambda V \mathbf{n}\!\cdot\!\mathbf{S}/\hbar$.
This is similar to the detection scheme 
presented in~\cite{Levitov1996} for charges transferred.
A fundamental complication in comparison with the charge FCS 
is that in our case the phase shift depends on spin, so that even the
minimal feedback influences the statistics 
of the outcomes of following spin detectors. 
We introduce dimensionless variables $N = 2\int dtQ/\hbar\lambda$, 
$\phi = \lambda V/2$. 
Then $N$ varies by one every time a spin 1/2 crosses the detector. 
With reference to Eqs.\eqref{eq:response},\eqref{eq:fluctuation}, we have 
$m=LC^2\to 0$, $\Omega^2=1/LC\to \infty$, 
$m\Omega^2\to C$, 
$\gamma\to RC^2$, 
$E=4C/\lambda^2$, 
with $L,$, $R$ and $C$ inductance (assumed negligible), 
resistance and capacitance of the circuit. 
\section{The setup considered}\label{sec:setup}
\begin{figure}[htb!]
\includegraphics[width=0.3\textwidth]{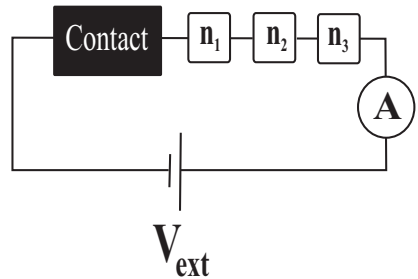}
\caption{\label{fig:setup}The setup considered, in the case of three spin 
detectors and one charge detector.}
\end{figure}

We consider a system composed of two metallic, 
unpolarized leads, connected through a 
coherent conductor, characterized by a set of transmission probabilities 
$T_n$, where $n$ identifies transmission channels. 
A negative bias voltage $V$ is applied to the left lead. 
At  the right of the conductor there are several spin detectors, labelled 
from left to right by $a=1,2\cdots$, and a current detector, denoted by $a=0$. 
The counting fields will be then $\phi_a$, with $a=0,1,\cdots$. 
Since charge and spin currents commute, 
the current detector can be positioned at any point 
along the chain of detectors, 
without influencing the statistics of the outcomes. 
The setup is depicted in Fig.~\ref{fig:setup}. 
We require that the coherent conductor is non-polarizing. 
Thus, the average spin current is zero. 
However, there are spin fluctuations, which are revealed by measuring 
noise and higher order correlators (or cumulants). 
From the symmetry with respect to reversal of spin, 
we can predict a priori that all 
odd cumulants are zero. 
We shall concentrate on a situation where 
there are three spin detectors. 
This is because, 
as anticipated in section \ref{sec:background}, 
the current is unpolarized and 
one needs at least three detectors 
monitoring non-commuting quantities in order to see 
non-trivial consequences of the detectors' feedback on the system. 
The feedback consists in the wave-function 
picking up an Aharonov-Casher phase while 
traversing each detector. 
\section{Results}\label{sec:technique}
The technique we  use is an extension of the
scattering theory of charge FCS. 
This theory~\cite{Levitov1993,Levitov1996,Belzig2001}
expresses FCS in terms of a phase factor $e^{i\chi}$ acquired
by scattering waves upon traversing the charge detector.

Since we do not consider energy-resolved measurements, 
the phase factor does not depend on the channel, and 
the approach works for a multi-channel conductor as well as
for a single-channel one.
The phase factor $e^{i\chi}$ can be seen as resulting from a gauge transform, 
to be applied to the (known) Green function of the right lead, 
that removes the coupling term~\cite{Levitov1996,Belzig2001} 
$\hat H_{int}=-\frac{\hbar}{e} \Hat\chi \hat I $ . 
For the case of the spin detectors, 
the gauge transform introduces a phase factor which is a unitary matrix in spin space. 
Namely, 
the gauge transform generated by spin detector $a$ is 
$e^{i \phi_{a} \mathbf{n}_{a}\cdot\bs{\tau}}$.
In this matrix, $\boldsymbol{\tau}$ is a pseudovector of $2\times2$
Pauli matrices, and $\mathbf{n}_a$ is the direction along which detector $a$ detects 
spin current. 
The Keldysh Green function of the lead is
\[\check{G}_{l}(E) =   
\begin{pmatrix}1-2f_l&-2f_l\\-2(1-f_l)&2f_l-1
\end{pmatrix}\;,\]
where $l\in\{L,R\}$ denotes the left or right lead, and 
$f_l$ is the corresponding Fermi occupation number 
at energy $E$ and chemical potential $\mu_l$. 
The elements of the matrix are actually in their turn a matrix in spin space. 
Since the leads are assumed to be unpolarized, they are simply the identity. 
The matrix current is given by \cite{Belzig2002}
\begin{equation}
\check{I}(\chi,\phi) = 
\frac{e^2}{2\pi\hbar}
\sum_{n} 
\frac{{T}_{n}\left[\check{G}_{L}
,\check{\tilde{G}}_{R}\right]}
{1+
{T}_{n} \left(\left\{ 
\check{G}_{L},\check{\tilde{G}}_{R}\right\}
-2\right)/4}\;,
\end{equation}
from which it follows that the quantum generating function is 
\[\mathcal{F} = \frac{e^2}{2\pi\hbar}
\sum_{n} \int dE 
\log{\left(1+{T}_{n} \left(\left\{ 
\check{G}_{L},\check{\tilde{G}}_{R}\right\}
-2\right)/4\right)}\;.\]
Here $\left[...\right](\{...\})$ denote (anti)commutator of 
two matrices, and $\check{\tilde{G}}_{R}$ is the transformed 
matrix 
\[\check{\tilde{G}}_{R} =  e^{i\bar{\chi}}  
\prod^{\rightarrow}_a e^{i\bar{\phi}_{a}\mathbf{n}_a\cdot \boldsymbol{\tau}}
\check{G}_{R}
\prod^{\leftarrow}_a e^{-i\bar{\phi}_{a}\mathbf{n}_a\cdot \boldsymbol{\tau}}
e^{-i\bar{\chi}}\;,\]
where $\bar{\chi}= \diag(\chi^+,\chi^-)$, $\bar{\phi} =\diag{(\phi^+,\phi^-)}$ 
are matrices in Keldysh space. 
After substituting the expression for $\check{G}_{R}$, we obtain  
\[\check{\tilde{G}}_{R} =  
\begin{pmatrix}
1-2f_R&-2f_R e^{i\chi}\mathcal{M}\\
-2(1-f_R) e^{-i\chi} \mathcal{M}^\dagger&2f_R-1
\end{pmatrix}
\;,\]
where 
\begin{equation}
{\cal M} \equiv \prod^{\rightarrow}_a
e^{i(\phi^+_{a}/2)\mathbf{n}_a\cdot \boldsymbol{\tau}}
\prod^{\leftarrow}_a
e^{-i(\phi^-_{a}/2)\mathbf{n}_a\cdot \boldsymbol{\tau}} \;.
\label{eq:matrix}
\end{equation}
is a matrix in spin space. 
We notice that 
\emph{i}) the charge fields come only in the combination 
$\chi = \chi^+-\chi^-$, and 
\emph{ii})
the phase factors
$e^{i \chi}$ in the expression for charge FCS
are replaced by $e^{i \chi}{\cal M}$ to give the FCS of charge
and spin counts after taking trace over spin. If we also notice
that ${\cal M}$ is a $(2\times 2)$ matrix with eigenvalues 
$e^{\pm i \alpha}$, we arrive to 
\begin{equation}
	\mathcal{F}(\chi,\{\phi^+_{a}\},\{\phi^-_{a}\}) 
=\frac{1}{2} \sum_{\pm} \mathcal{F}_{c}(\chi \pm \alpha)\;,
\label{eq:main_relation}
\end{equation}
where $\mathcal{F}_c(\chi)$ is the generating function for charge counting. 
The $\alpha$ is given by 
\begin{equation}\label{eq:matrixeigen}
\cos{\alpha}=\frac{1}{2}tr {\cal M}
\end{equation}. 

The explicit expression for the system considered here, 
in terms of the transmission 
probabilities through the contact and the applied bias is, 
at zero temperature, 
\begin{equation}
\mathcal{F} 
=\int_0^\tau\frac{dt}{\tau_V} \sum_n\log{\left[R_n^2 + T_n^2 e^{2i\chi} + 2 R_n T_n e^{i\chi}\cos{\alpha}
\right]}\;,
\label{eq:main_relation2}
\end{equation}
with $R_n\equiv 1-T_n$, $\tau_V \equiv 2\pi\hbar/eV$. 
The interpretation is quite straightforward: 
electrons coming through different channels 
behave independently, which is revealed by the 
fact that the generating function splits into a sum; 
each channel can accommodate two electrons in a spin-singlet configuration; 
with probability $R_n^2$ none of the 
two electrons passes the junction, and there is no contribution 
to the charge counting nor to the spin one; 
with probability $T_n^2$ both electrons come through the conductor. 
This gives a contribution of 
two elementary charges transferred (factor $e^{2i\chi}$), but no 
spin transfer. Finally, with probability 
$p_n=2R_n T_n$, exactly one of the two electrons is transferred. 
This gives a contribution to the charge and to the spin counting.
\section{Projection Postulate}\label{sec:PP}
We demonstrate that a different FCS is predicted by 
using a different approach, 
namely a na\"{\i}ve application of the 
projection postulate, 
consisting in avoiding the description of the measurement and 
applying the projection to the system measured. 
We shall denote this procedure with PP for brevity. 
This approach predicts a parameter $\alpha_{PP}$ 
which does not depend on $\Phi$. 
Let us give the details of such a derivation: 
When an unpolarized electron arrives to the first detector, the probability of 
the outcome $\sigma_1 = \pm 1$ is $P_1(\sigma_1)=1/2$. 
The conditional probability that the second detector gives 
$\sigma_2$, given that the first read $\sigma_1$ is 
$P_2(\sigma_2|\sigma_1) 
= (1+\sigma_1 \sigma_2 \mathbf{n}_1\cdot\mathbf{n}_2)/2$. 
This is because after the first detection 
the spin of the electron is assumed to have collapsed 
along $\pm \mathbf{n}_1$. The same happens after the second detection. 
Consequently, the conditional probability 
that a third detector reads $\sigma_3$, given that 
the first read $\sigma_1$ and the second $\sigma_2$, 
depends only on the latter outcome 
$P_3(\sigma_3|\sigma_2,\sigma_1) 
= (1+\sigma_2 \sigma_3 \mathbf{n}_2\cdot\mathbf{n}_3)/2$. 
The process is in a sense a Markovian one.
The total joint probability for each electron transfer with 
an arbitrarily long chain of detectors is 
\[P(\{\sigma\}) = \frac{1}{2} \prod_{a=1}^{K-1} 
\frac{1\!+\!\sigma_{a} \sigma_{a+1} 
\mathbf{n}_a\!\cdot\!\mathbf{n}_{a+1}}{2} \;,\]
and the corresponding generating function for the setup considered here 
is given by Eq.~\eqref{eq:main_relation2} with 
\begin{align}
\cos{\alpha_{PP}} = \sum_{\{\sigma\}} \cos{\left[\sum_a\sigma_a \phi_a\right]}	
\frac{1}{2} \prod_{a=1}^{K-1} 
\frac{1\!+\!\sigma_{a} \sigma_{a+1} 
\mathbf{n}_a\!\cdot\!\mathbf{n}_{a+1}}{2} \;.
\end{align}
\section{Comparison of the two approaches 
for one and two spin detectors}\label{sec:comp12}
Now, let us go back to Eqs.~\eqref{eq:matrix},\eqref{eq:matrixeigen} 
and compare the two approaches for some simple cases. 
For the case of one or two detectors in series, 
the eigenvalues $e^{\pm i\alpha}$
are not affected by the order of matrix multiplication in (\ref{eq:matrix})
and depend on differences of spin counting fields 
$\phi_a \equiv \phi^+_{a} -\phi^-_{a}$ only 
(in fact they coincide with the value $e^{\pm i \alpha_{PP}}$). 
This implies that the FCS definition (\ref{eq:FCS_definition}) can be readily
interpreted in classical terms: it is a generating function 
for probability distribution of a certain number of spin counts 
$\sigma_a$ in each detector,
\begin{equation}
P(\{\sigma_a\}) = \int \prod_{a} d \phi_a 
e^{F(0,\{\phi_a\})} e^{- i\sum_a \sigma_a \phi_{a}}
\label{eq:probability} 
\end{equation}
For a single detector, the spin FCS is very simple: 
it corresponds to independent transfers of 
two sorts of electrons, with spins "up" and "down"
with respect to the quantization axis. 
The cumulants of the 
 spin (charge) transferred
are given by the derivatives of $F$ with respect to $\phi_1$ ($\chi$),
at $\chi=\phi_1=0$. In this case $\alpha=\phi_1$. 
From this and relation (\ref{eq:main_relation}), 
we conclude that all odd cumulants of spin current
are 0, as anticipated, and all even cumulants 
coincide with the charge cumulants. 

For two spin detectors, 
with $\mathbf{n}_1\cdot\mathbf{n}_2=\cos{\theta}$, 
we obtain 
$\cos\alpha=\cos\phi_1\cos\phi_2-\sin\phi_1\sin\phi_2\cos\theta$. 
Since there is no dependence on $\Phi_a$, 
the quantum generating function has an immediate interpretation; 
we consider the case 
when the read-out of the charge is not exploited ($\chi=0$). 
Then 
\begin{equation}
\mathcal{Z}(\phi) = \prod_n\left[q_n + p_n\cos{\alpha}\right]^M\!,
\end{equation}
where $p_n=2 R_n T_n$ is the probability that, in two attempts of transmitting 
one electron over a spin degenerate channel $n$, 
exactly one is transmitted and $q_n=1-p_n$.
This result coincides with what one would obtain 
from the Projection Postulate. 

We discuss in detail the probability distribution. 
By performing the Fourier transform, we find the probability of 
detecting a spin $\sigma_1$ in direction $\mathbf{n}_1$ and 
$\sigma_2$ in direction $\mathbf{n}_2$: 
\begin{equation}
P(\sigma_1,\sigma_2) = 
{\sum_{\sigma_1^{(n)}}}'{\sum_{\sigma_2^{(n)}}}' 
\prod_n P_n(\sigma_1^{(n)},\sigma_2^{(n)})\;,
\end{equation}
where the prime in the sum means that it is restricted to 
$\sum_n \sigma_a^{(n)} =\sigma_a$, 
and the probability for each channel $n$ is 
\begin{align}
\nonumber
P_n(\sigma_1,\sigma_2)\!=&\ \sum_k P_{tr}(k|N) P_\up((k+\sigma_1)/2|k) \\ 
\nonumber&\times
\sum_l P_\up((k+\sigma_2+2l)/4|(k+\sigma_1)/2=\up) \\ 
\label{2detprob}&\quad\times
P_\up((k+\sigma_2-2l)/4|(k-\sigma_1)/2=\dwn)
\end{align}
where 
\begin{align}
P_{tr}(k|N) =&\ \binom{N}{k}p_n^k q_n^{N-k}\;,\\
P_\up(l|k) =&\ \frac{1}{2^k} \binom{k}{l}\;,\\
P_\up(l|k=\up) =&\ \binom{k}{l}
[\cos^2{(\theta/2)}]^{l} [\sin^2{(\theta/2)}]^{k-l}\;,\\
P_\up(l|k=\dwn)=&\ 
\binom{k}{l} 
[\cos^2{(\theta/2)}]^{k-l} [\sin^2{(\theta/2)}]^{l}
\end{align}
The sums are over all values for which the binomials make sense 
(no negative nor half-integer values). Thus $k,l,\sigma_1, \sigma_2$  
have the same parity. 
$P$ can be interpreted as follows: since the current is unpolarized, 
we can think of it as carried by pairs of electrons in singlet 
configuration. 
Then, there is a successful attempt to transfer spin when exactly one 
of the two electrons is transmitted. This gives $P_{tr}(k|N)$, 
the probability of transferring $k$ spins 
over $N$ attempts ($p_n$ probability of success for a single attempt) 
through channel $n$; 
the second binomial comes from the ways one can pick 
$N_{1\up}=(k+\sigma_1)/2$ spins up out of 
$k$ spins, with probability $1/2$ (we recall that the incoming 
electrons are unpolarized); 
the third term comes from the fact that, 
given that $N_{1\up}=(k+\sigma_1)/2$ 
spins up according to the first detector are passed to the second one, 
the latter will measure $(k+\sigma_2+2l)/4$ of these as spins up 
(the probability of agreement 
between detectors being $p_{ag}=\cos^2(\theta/2)$), 
and the rest as spins down; 
analogously, the latter term comes from the fact that given 
$N_{1\dwn}=(k-\sigma_1)/2$ 
spins down along direction $\mathbf{n}_1$ have been detected, 
$(k+\sigma_2-2l)/4$ of them will be detected from 
the second detector as spins up, 
while the remaining ones will be detected as down.  

When the two detectors have parallel orientation ($\theta=0$), 
the second sum in Eq.~\eqref{2detprob} is nonzero 
only if $\sigma_1=\sigma_2$, giving
\[P(\sigma_1,\sigma_2)\!= \sum_k P_{tr}(k|N) P_\up((k+\sigma_1)/2|k)
\delta_{\sigma_1,\sigma_2}, 
\]
i.e.~there is perfect correlation, as is to be expected. 
When the two detectors have orthogonal orientation ($\theta=\pi/2$), 
it is possible to perform analytically the sum over $m$: 
\[P(\sigma_1,\sigma_2)\!= 
\sum_k P(k|N) P_\up((k+\sigma_1)/2|k)P_\up((k+\sigma_2)/2|k)
\]
i.e.~the outcomes are independent, given that $k$ successful spin 
transfers happened. 
\section{Comparison of the two approaches 
for three spin detectors}\label{sec:comp3}
For the case of three detectors, we have 
\begin{align}\nonumber
&\cos\alpha =\cos{\alpha_{PP}} -\sin{\theta_{12}}\sin{\theta_{23}} 
\sin{(\Phi_2-\Phi_2^{(0)})} \sin{\phi}_3 \sin{\phi}_1
, \nonumber \\
&\cos\alpha_{PP} = \cos{\phi}_1 \cos{\phi}_2 \cos{\phi}_3 +\mbox{ }\nonumber \\
&-\cos{\theta_{12}}\sin{\phi}_1 \sin{\phi}_2  \cos{\phi}_3 
-\cos{\theta_{23}}\sin{\phi}_2 \sin{\phi}_3 \cos{\phi}_1 
-\nonumber \\
& \cos{\theta_{12}}\cos{\theta_{23}}
\sin{\phi}_3 \sin{\phi}_1 \cos{\phi}_2 
. 
\label{avsapp}
\end{align}
Here $\theta_{jk}=\arccos{{\mathbf n}_j\!\cdot\!{\mathbf n}_k}$ are the angles 
between the polarizations ($\mathbf{n}$) of detectors $j$ and $k$, 
and $\cos{\Phi_2^{(0)}}= ({\mathbf n}_1 \times{\mathbf n}_2)\!\cdot\!{\mathbf n}_3/
\sin{\theta_{12}}\sin{\theta_{23}}$, 
$\sin{\Phi_2^{(0)}}= ({\mathbf n}_1 \!\times\!{\mathbf n}_2)\cdot 
({\mathbf n}_2\times\!{\mathbf n}_3)/
\sin{\theta_{12}}\sin{\theta_{23}}$. 
As before $\cos\alpha_{PP}$ is the part corresponding 
to the Projection Postulate. 
We notice that when two consecutive detectors are parallel or antiparallel, 
then 
$\alpha_{PP}=\alpha$. This is because the same measurement is repeated twice, 
and thus we fall back to the case of two detectors. 

In general, however, $\cos\alpha$ depends on $\Phi_2$, 
and thus one needs to account 
for the dynamics of the second detector 
in order to get the probability distribution 
for the spin counts. We recall that the corresponding detector's action is 
$S[\Phi_2]=\int dt \frac{1}{2}\left[\tau_c 
\dot{\Phi}_2(t)^2 - \Phi_2(t)^2/\tau_c \langle\Phi_2^2 \rangle\right]$, 
with $\tau_c$ coherence time and $\langle\Phi_2^2 \rangle$ 
fluctuations of $\Phi_2$. 

We have calculated the second cumulants or cross-correlators: 
we found that they differ 
from the ones obtained by using PP only by small terms. 
The correlator between first and third detector's readings is: 
\begin{align}
&\langle\!\langle \sigma_1 \sigma_3\rangle\!\rangle \!=\! 
\langle\!\langle N^2\rangle\!\rangle 
\left[
C + 
\left(
\cos{\theta_{13}}- 
C
\right)
e^{-\langle \Phi_2^2\rangle/2}
\right], 
\end{align}
where 
$C \equiv \cos{\theta_{12}}\cos{\theta_{23}}$,
and the first term is 
the PP result. 
The second term, as expected, has a typical signature of interference effects:
it is suppressed exponentially if the variance of the corresponding 
Aharonov-Casher phase $\langle\Phi_2^2\rangle \gg 1$. Since 
$\Phi_{AC}$ is inversely proportional to $\hbar$, this is the 
classical limit. In this limit, the result coincides with the PP.

However, fourth cumulants show a large deviations from the PP result. 
Namely: 
\begin{align}\label{eq:QMpredictions}
\langle\!\langle \sigma_1^2 \sigma_3^2\rangle\!\rangle &\!=\! 
\langle\!\langle 
\sigma_1^2 \sigma_3^2\rangle\!\rangle_{\raisebox{-1mm}{$_{\!\!PP}$}}
+8\frac{\tau_c}{\tau}
A
\langle\!\langle N^2\rangle\!\rangle^2 ,
\end{align}
where \mbox{$A\equiv \sin^2{\theta_{12}}\sin^2{\theta_{23}}$,} 
and the PP result is expressed in terms of charge cumulants as 
\begin{equation*}
\langle\!\langle 
\sigma_1^2 \sigma_3^2\rangle\!\rangle_{\raisebox{-1mm}{$_{\!PP}$}}\!\!=
\frac{1}{3}\left[(1+2C^2)\langle\!\langle N^4\rangle\!\rangle 
+ 2 (1-C^2) \langle\!\langle N^2\rangle\!\rangle\right] .
\end{equation*} 
This deviation results from correlations of $\Phi_2$ at time scale $\tau_c$.
To estimate the result, we notice that the charge cumulants are of the order
of $\tau/\tau_{el}$, $\tau_{el}$ being the average time between electron
transfers. It is easy to fulfill the condition 
$\tau_{el} \ll \tau_{c} \ll \tau$, and in this case 
$\langle\!\langle \sigma_1^2 \sigma_3^2\rangle\!\rangle$ 
is much larger than PP result.   

It is interesting to study further the probability 
distribution which gives rise 
to such anomalously large fourth-order cumulants. 
This we shall do in the next section. 
\section{A particular case}\label{sec:case}
We discuss for definiteness the case of three detectors oriented 
along three orthogonal directions forming a 
right-handed basis. 
This implies $\Phi_2^{(0)}=0$. 
We concentrate on the joint probability distribution 
for the outcomes of the first and the third detector, 
irrespectively of the reading of the second detector. 
We consider the "classical" limit $\langle \Phi_2^2 \rangle \to \infty$. 
Then, 
the generating function for the probability $P(\sigma_1,\sigma_3)$ for 
counting $\sigma_1, \sigma_3$ spins in the detectors is: 
\begin{align}
\nonumber
Z(\phi_1,\phi_3) =&\ \int d\Phi_{2,i} d\Phi_{2,f} 
\int_{\Phi_{2,i}}^{\Phi_{2,f}} {\mathcal{D}}\Phi_2(t) 
\exp\Biggl\{\int_0^\tau dt\\
&\left[-\frac{\tau_c}{2}\dot{\Phi}_2^2 + \frac{1}{\tau_V}
\sum_n\ln{\left[q_n+p_n\cos{\alpha(\phi,\Phi_2)}\right]}\right] \Biggr\}
\end{align}
where
\begin{align}\nonumber
&\cos\alpha(\phi,\Phi_2) =\cos{\alpha_{PP}} -
\sin{\phi}_3 \sin{\phi}_1 
\cos\Phi_2
, \nonumber \\
&\cos\alpha_{PP} = \cos{\phi}_1  \cos{\phi}_3\;. 
\label{eq:avsapp}
\end{align}
We have a path-integral over imaginary time. 
We exploit the quantum mechanical technique 
and re-express the path-integral in terms 
of amplitudes: 
\[Z(\phi)=\int d\Phi_{2,i} d\Phi_{2,f} 
\langle\Phi_{2f};t=i\tau,\phi|\Phi_{2i}\;t=0,\phi\rangle\;.\] 
Here the counting fields $\phi$ are parameters, and the time evolution 
of the variable $\Phi_2$ 
is dictated by 
$|\Phi_2;t,\phi\rangle = e^{-i\Hat{H}(\phi)t}|\Phi_2;0,\phi\rangle$, 
with the Hamiltonian
\begin{equation}
\Hat{H}(\phi) = -\frac{1}{2\tau_c}\frac{\partial^2}{\partial \Phi_2^2} 
- \frac{1}{\tau_V} 
\sum_n\ln{\left[q_n+p_n\cos{\alpha(\phi,\Phi_2)}\right]}
\;.
\end{equation}
Then,
for large values of $\tau$, the path-integral can be approximated
\begin{equation}
Z(\phi) \simeq e^{-E_0(\phi)\tau}\;,
\end{equation}
where $E_0(\phi)$ is the ground state energy of the Hamiltonian.

The next step is to find an explicit expression for the probability. 
We recall that the probability to have detectors 1 and 3 measure average 
spin currents $I_1=\sigma_1/\tau$, $I_3=\sigma_3/\tau$ is related to $Z(\phi)$ 
through 
\[P(I_1,I_3) = \int \frac{d\phi_1}{2\pi}\frac{d\phi_3}{2\pi} Z(\phi) 
e^{-i\tau (\phi_1 I_1+\phi_3 I_3)}\;.\]
Since we are in the large $\tau$ limit, 
we can evaluate the integrals in the saddle-point approximation, 
and obtain
\[P(I_1,I_3) \propto \exp{
\left[-E_0(\phi^*)-i\tau (\phi^*_1 I_1+\phi^*_3 I_3)\right]}\;,\]
where 
$\phi^*_a$ satisfy the saddle point condition
\begin{subequations}\label{eq:sp}
\begin{align}
\left.\frac{\partial E_0}{\partial \phi_1}
\right|_{\phi_1^*,\phi_3^*}+i I_1 =& 0 \;,\\
\left.\frac{\partial E_0}{\partial \phi_3}
\right|_{\phi_1^*,\phi_3^*}+i I_3 =& 0 \;
\end{align}
\end{subequations}
Assuming that the solutions are much smaller than 1, $\phi^*_a \ll 1$, 
we have that the Hamiltonian can be rewritten, 
including terms up to second order in $\phi$, as 
\begin{equation}
\Hat{H}(\phi) = \left[-\frac{1}{2\tau_c}\frac{\partial^2}{\partial \Phi_2^2} 
+ \frac{1}{2\tau_S} 
\left(\phi_1^2+\phi_3^2+2\phi_1\phi_3 \cos{\Phi_2}\right)\right]
\;,
\end{equation}
where we introduced the average time between spin transfers, 
$\tau_S=\tau_V/\sum_n p_n$. 
We recognize the Hamiltonian for the Mathieu equation 
\[H_M = -\frac{\partial^2}{\partial v^2} + 2 q \cos{(2v)}\;.\] 
Thus the ground state energy depends on the lowest Mathieu 
characteristic function $a_0(q)$, 
with the coupling strength given by $q=4(\tau_c/\tau_S)\phi_1\phi_3$. 
Namely, 
\[E_0(\phi) = a_0(q)/8\tau_c + (\phi_1^2+\phi_3^2)/2\tau_S\;.\] 
The saddle-point equations \eqref{eq:sp} can 
then be combined to give a trascendent 
equation for $q$, from which one expresses $\phi_a^*$, 
which are purely imaginary, according to 
\begin{subequations}\label{eq:spsol}
\begin{align}
i\phi_1^* =&\ \tau_S\frac{I_1- (I_3/2) a'_0(q^*)}{1-a'_0(q^*)^2/4}\;,\\
i\phi_3^* =&\ \tau_S\frac{I_3- (I_1/2) a'_0(q^*)}{1-a'_0(q^*)^2/4}\;.
\end{align}
\end{subequations}
Here, $q^*$ is the solution to the equation 
\begin{equation}\label{eq:selfcons}
\frac{q}{4} = -\frac{(\nu_1+\nu_3)^2}{\left[2+a'_0(q)\right]^2} 
+\frac{(\nu_1-\nu_3)^2}{\left[2-a'_0(q)\right]^2} 
\;,
\end{equation}
where we introduced dimensionless currents 
$\nu_a \equiv \sqrt{\tau_c \tau_S} I_a$.
Eqs. \eqref{eq:spsol},\eqref{eq:selfcons} are valid in the limit 
$\tau_S I_a \ll 1$, i.e. $\nu_a \ll \sqrt{\tau_c/\tau_S}$.  

Finally, we have that the probability distribution is 
\begin{align}\nonumber
\log P(I_1,I_3)
\propto& -a_0(q^*)/8 \\
\nonumber
&- (\nu_1+\nu_3)^2 (1+a'_0(q^*))/
\left[2+a'_0(q^*)\right]^2 \\&
\label{eq:prob}
- (\nu_1-\nu_3)^2 (1-a'_0(q^*))/
\left[2-a'_0(q^*)\right]^2\;. 
\end{align}

This probability distribution is to be compared 
with the one predicted by applying the PP. 
The latter is, in the same regime $\tau_S I_a \ll 1$, 
the independent combination of two gaussians: 
\begin{equation}\label{eq:ppprob} 
\log P_{PP}(I_1,I_3)\propto -(\nu_1^2+\nu_3^2)/2\;,
\end{equation}
the proportionality constant ($\tau/\tau_c$) 
being the same. 

In the limit $\tau_c \ll \tau_S$, 
we have that Eqs. \eqref{eq:prob} and \eqref{eq:ppprob} coincide. 
However, by taking into account 
that the detectors have a finite decoherence time $\tau_c$, 
and that the time between spin transfers 
$\tau_S$ can be much smaller than $\tau_c$, 
we find that the probability distribution deviates 
sensibly from Eq.\eqref{eq:ppprob}. 
This deviation is larger 
in the regime $1\ll|\nu_1|\simeq |\nu_3| \ll\sqrt{\tau_c/\tau_S}$. 
when both (dimensionless) currents 
are comparable in module and large with respect to 1. 

When $\nu_1\gg 1$, we find that 
\begin{equation}\label{eq:fcspred}
\log{P} \propto -\nu_1^2/2 + f(\nu_3/\nu_1)\;,
\end{equation}
with the scaling function $f(x)$ defined by 
\[f(x)=-\frac{a_0(q_0(x))}{8}
+\frac{1}{4}x q_0(x)\]
where the condition 
$\left.\frac{\partial a_0}{\partial q}\right|_{q=q_0}=2x$ defines $q_0(x)$. 
In particular, $f(x)$ diverges at $x=1$ according to 
$f(x)\simeq -1/16(1-x)$. 
\begin{figure}[t!]
\includegraphics[width=0.4\textwidth]{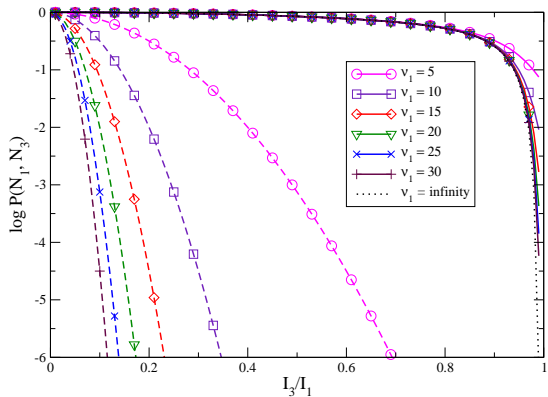}
\caption{\label{fig:comp}
The log of probability as a function of $\nu_3/\nu_1$ 
for different values of $\nu_1$ 
for the configuration studied in the text. 
All the curves have been shifted by $\nu_1^2/2$. 
The upper curves correspond to the result of the FCS approach, 
and the lower ones 
to the PP. The black dotted curve 
is the limiting scaling curve discussed in the text.}
\end{figure}
In Fig.~\ref{fig:comp} we draw the logarithm of probability 
as a function of $\nu_3/\nu_1$ for several values of $\nu_1$, 
and compare with the probability 
predicted by making use of PP. 
\section{Conclusions}\label{sec:conclusions}
We have discussed the Full Counting Statistics of 
non-commuting variables. As a concrete example, 
we focused on 
spin counts in a two terminal device 
with non-ferromagnetic leads connected 
through a non-polarizing coherent conductor. 
We have provided a formula connecting the FCS of spins to the one of charge. 
We have seen that it is crucial 
to have a coherent conductor with finite transparency 
connecting the two leads. 
This is because electrons transmitted 
through the same channel are in a spin singlet, and thus 
contribute no net spin transfer nor spin fluctuations. 
However, if the transmission probability through channel $n$ 
is finite ($0<T_n<1$), 
then there is a non-zero probability $p_n= 2 (1-T_n) T_n$ that 
exactly one electron out of a singlet pair is transmitted, 
and this contributes to spin fluctuations. 

Another interesting conclusion which we can draw from this work is that, 
when measuring non-commuting quantities with subsequent detectors, 
one should take into account 
the quantum dynamics of the detectors themselves. 
This is because the decoherence time 
for the detectors, $\tau_c$ can be larger than the average time between 
two subsequent counts, $\tau_S$. Thus if one would 
put by hand the off-diagonal elements of the detectors' density matrix to zero 
 after each count, which amounts to applying the projection postulate, one 
would obtain the wrong result. We have shown that, 
in the system considered here, such a deviation from the 
na\"ive application of the projection postulate is revealed 
by the fourth correlator of spin counts.  
\acknowledgments
We acknowledge the financial support provided through the European
Community's Research Training Networks Programme under contract
HPRN-CT-2002-00302, Spintronics. 

\end{document}